\documentclass[10pt,letterpaper]{article}
\usepackage[top=0.85in,left=1in,footskip=0.75in,marginparwidth=2in]{geometry}

\usepackage[utf8]{inputenc}

\usepackage{cite}

\usepackage{nameref,hyperref}
\usepackage{amsfonts, amsmath, amssymb,latexsym,amsthm, mathrsfs, enumerate}

\usepackage{microtype}
\DisableLigatures[f]{encoding = *, family = * }

\usepackage{changepage}

\usepackage[aboveskip=1pt,labelfont=bf,labelsep=period,singlelinecheck=off]{caption}

\makeatletter
\renewcommand{\@biblabel}[1]{\quad#1.}
\makeatother

\usepackage{lastpage,fancyhdr,graphicx}
\usepackage{epstopdf}
\fancyheadoffset[L]{2.25in}
\fancyfootoffset[L]{2.25in}

\begin{document}
\vspace*{0.35in}

\begin{flushleft}
{\Large
\textbf\newline{Slow light mediated by mode topological transitions in hyperbolic waveguides}
}
\newline
\\
Pilar Pujol-Closa\textsuperscript{1},
Jordi Gomis-Bresco\textsuperscript{1},
Samyobrata Mukherjee\textsuperscript{1},
J. Sebasti\'an G\'omez D\'iaz\textsuperscript{2},
Lluis Torner\textsuperscript{1,3},
David Artigas\textsuperscript{1,3,*}
\\
\bigskip
\bf{1} ICFO - Institut de Ciencies Fotoniques, The Barcelona Institute of Science and Technology, Castelldefels (Barcelona), Spain
\\
\bf{2} Department of Electrical and Computer Engineering, University of California Davis, Davis, USA
\\
\bf{3} Department of Signal Theory and Communications,
Universitat Polit\`ecnica de Catalunya, Barcelona, Spain
\\
\bigskip
* Corresponding author: david.artigas@icfo.eu

\end{flushleft}
\noindent\rule[0.5ex]{\linewidth}{1pt}

\section*{Abstract}
We show that symmetric planar waveguides made of a film composed of a type II hyperbolic metamaterial, where the optical axis (OA) lies parallel to the waveguide interfaces, result in a series of topological transitions in the dispersion diagram as the film electrical thickness increases. The transitions are mediated by elliptical mode branches, which, as soon as they grow from cutoff, coalesce along the OA with anomalously ordered hyperbolic mode branches, resulting in a saddle point. When the electrical thickness of the film increases further, the merged branch starts a transition to hyperbolic normally ordered modes with propagation direction orthogonal to the OA. In this process, the saddle point is transformed into a branch point where a new branch of Ghost waves appears and slow light is observed for a broad range of thicknesses.

\noindent\rule[0.5ex]{\linewidth}{1pt}

\section*{}

\noindent Metamaterials were first theoretically hypothesized in the late 1960s by Veselago \cite{Veselago}, reaching maturity in the last two decades thanks to new proposals that made fabrication feasible \cite{Pendry, Smith}. Among them, Hyperbolic Metamaterials (HMMs) are highly anisotropic media that have recently attracted significant attention due to their properties and application prospects.  Hyperlensing \cite{Jacob:06,Liu1686,Fang534,PhysRevB.86.121108} was one of the first proposed applications, which led to better nanolithography techniques \cite{doi:10.1063/1.2985898,doi:10.1021/acs.nanolett.6b04175}, and enhancement and control of directional spontaneous emission \cite{doi:10.1063/1.4824287}. They have been proposed in integrated optics \cite{Kildishev1232009,SmithNat} and as alternative materials in sensing, featuring an increased sensitivity and smaller device dimensions \cite{StrangiNatSens,PerfectAbsorber}. 

Planar waveguides with type II HMMs have been widely studied for the particular case in which propagation direction coincides with one of the  principal axis of the dielectric tensor only \cite{KildishevOL14}. For example, anomalous ordering \cite{Anomalous_Ordering}, cut-off and slow light near cut-off have been reported in these configurations \cite{Jiang:09,Alekseyev:06,Slow_light}, which also result in a high density of states and large Purcell factor \cite{PhysRevB.80.195106}. In the limit when the waveguide is extremely thin, i.e. resulting in hyperbolic metasurfaces, the structure features topological transitions by changing the material permittivities, from elliptical to hyperbolic dispersion \cite{Krishnamoorthy,SebastianTopo}, which are relevant to enhance spontaneous emission for high efficiency thermophotovoltaic conversion \cite{JacobTopo}. In this paper we extend dispersion analysis to arbitrary wave propagation directions in symmetric waveguides with a type II HMM as a film, where the optical axis (OA) lies parallel to the interfaces \cite{ZHOU2018145}. We show that, as the wave frequency increases, anomalously ordered hyperbolic modes with backward power flow and propagation close to the OA perform a series of topological transitions towards normally ordered hyperbolic modes, with positive power flow, propagating orthogonally to the OA. Interestingly, new phenomena as slow light and Ghost waves appear during the transition process. 

\begin{figure}[h]
\centering
\includegraphics[width=0.5\linewidth]{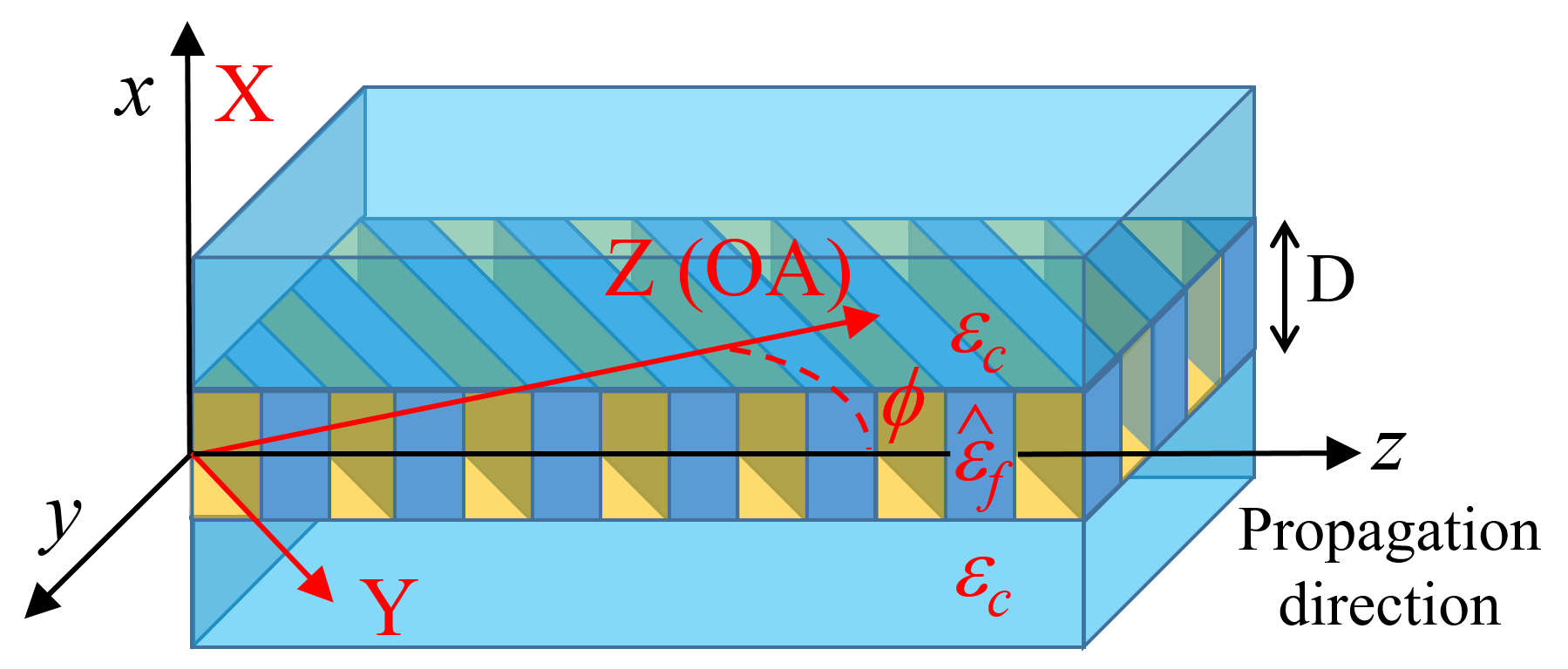}
\caption{Planar symmetric waveguide with a film thickness $D$ made of a type II hyperbolic metamaterial sandwiched between identical isotropic dielectric cladding and substrate. The film could be made of metal/dielectric (blue/yellow) multi-layers, with metamaterial reference axis $X, Y, Z$ (red color), so that $Z$ is the optic axis (OA) orthogonal to the multi-layer and parallel to the cladding-film-substrate interfaces. The dispersion properties of the supported modes are analyzed using an auxiliary coordinate system ($x, y, z$ in black color) which is rotated an angle $\phi$ with respect to the OA so that $z$ is along the wave propagation direction.}
\label{fig:conceptual}
\end{figure}

Figure \ref{fig:conceptual} shows the waveguiding system made of a  film (or core) with thickness $D$ and a uniaxial permittivity tensor $\hat{\varepsilon}_f$, which in the metamaterial reference system ($X, Y, Z$ in red) is diagonal with positive extraordinary and negative ordinary dielectric constants ($\varepsilon_Z=\varepsilon_{ef}>0$, $\varepsilon_{X,Y}=\varepsilon_{of}<0$). The substrate and cladding have identical dielectric constant $\varepsilon_c$. In this configuration, the OA lies parallel to the waveguide interfaces and along the $Z$ direction. An auxiliary coordinate system ($x,y,z$, in black), rotated an angle $\phi$ with respect to the OA, is employed to calculate the mode dispersion so that the propagation direction is along $z$ and determined by $\phi$. Electric field  components are locally defined with respect this coordinate system. In this notation the dispersion equations for modes with odd and even $E_y$ field component with respect the $x$ axis are respectively: 

\begin{center}
\begin{multline}
    \left(\gamma_{of} F \sinh{(k_0\gamma_{of} D)}+\gamma_c G \cosh{(k_0\gamma_{of} D})\right) \cdot 
    \left(|\varepsilon_{of}| \kappa_{ef} F \cos{(k_0\kappa_{ef} D)}-\varepsilon_c \gamma_c G \sin{(k_0\kappa_{ef} D)}\right) + \\ 
    \left(\varepsilon_c \gamma_{of}^2+|\varepsilon_{of}| \gamma_c^2\right)^2 \cosh{(k_0\gamma_{of}D)}\sin{(k_0\kappa_{ef}D)} = 0,
    \label{transodd}
\end{multline}
\end{center}
and
\begin{center}
\begin{multline}
    \left(\gamma_{of} F \cosh{(k_0\gamma_{of} D)}+\gamma_c G \sinh{(k_0\gamma_{of} D})\right) \cdot 
     \left(|\varepsilon_{of}| \kappa_e F \sin{(k_0\kappa_{ef} D)}+\varepsilon_c \gamma_c G \cos{(k_0\kappa_{ef} D)}\right) - \\ 
    \left(\varepsilon_c \gamma_{of}^2+|\varepsilon_{of}| \gamma_c^2\right)^2 \sinh{(k_0\gamma_{of}D)}\cos{(k_0\kappa_{ef}D)}= 0.
    \label{transeven}
\end{multline}
\end{center}
Where
\begin{equation}
\begin{aligned}
     \kappa_{ef}^2  = \varepsilon_{ef}+N_{eff}^2 & \left(\dfrac{\varepsilon_{ef}}{|\varepsilon_{of}|}\cos^2{\phi}-\sin^2{\phi}\right), \\
     \gamma_{of}^2  =  N_{eff}^2+|\varepsilon_{of}|,  
         & \qquad \gamma_c^2  = N_{eff}^2-\varepsilon_c, \\ 
     F  = \gamma_c^2 - \varepsilon_c \tan^2{\phi}, 
         & \qquad G  = \gamma_{of}^2 + |\varepsilon_{of}| \tan^2{\phi}.
\end{aligned}
\end{equation}
\noindent Here, $k_e$ and $k_o=i \gamma_{of}$ are the transverse wavevector component in the film for the extraordinary and ordinary waves, respectively. The decaying constant at the cladding and substrate is $\gamma_{c}$. $N_{eff}$ is the effective index of the mode, $k_0 = 2 \pi/\lambda$ is the free space wavenumber, and $\lambda$ the free space wavelength. In this notation, the momentum components in the metamaterial reference axes are given by $k_Y=N_{eff}k_0 \sin{\phi}$ and $k_Z=N_{eff}k_0 \cos{\phi}$. Then, by solving (\ref{transodd}) and (\ref{transeven}) or using the transfer matrix method described in Ref. \cite{ourPRA}, the effective index $N_{eff}$ in terms of the electrical thickness $D/\lambda$ and the propagation direction $\phi$ can be obtained. We investigate HMM waveguides versus their electrical thickness for a fixed wavelength. Alternative analysis based on varying the operation wavelength for a fixed film thickness is also possible, yielding to similar qualitative results. Without loss of generality, our simulations are performed for material permittivities $\varepsilon_c=3$, $\varepsilon_{ef}=4$ and $\varepsilon_{of}=-4$ and non-local effects are not considered \cite{Lavrinenko}. 

\begin{figure}[htbp]
\centering 
\includegraphics[width=0.5\linewidth]{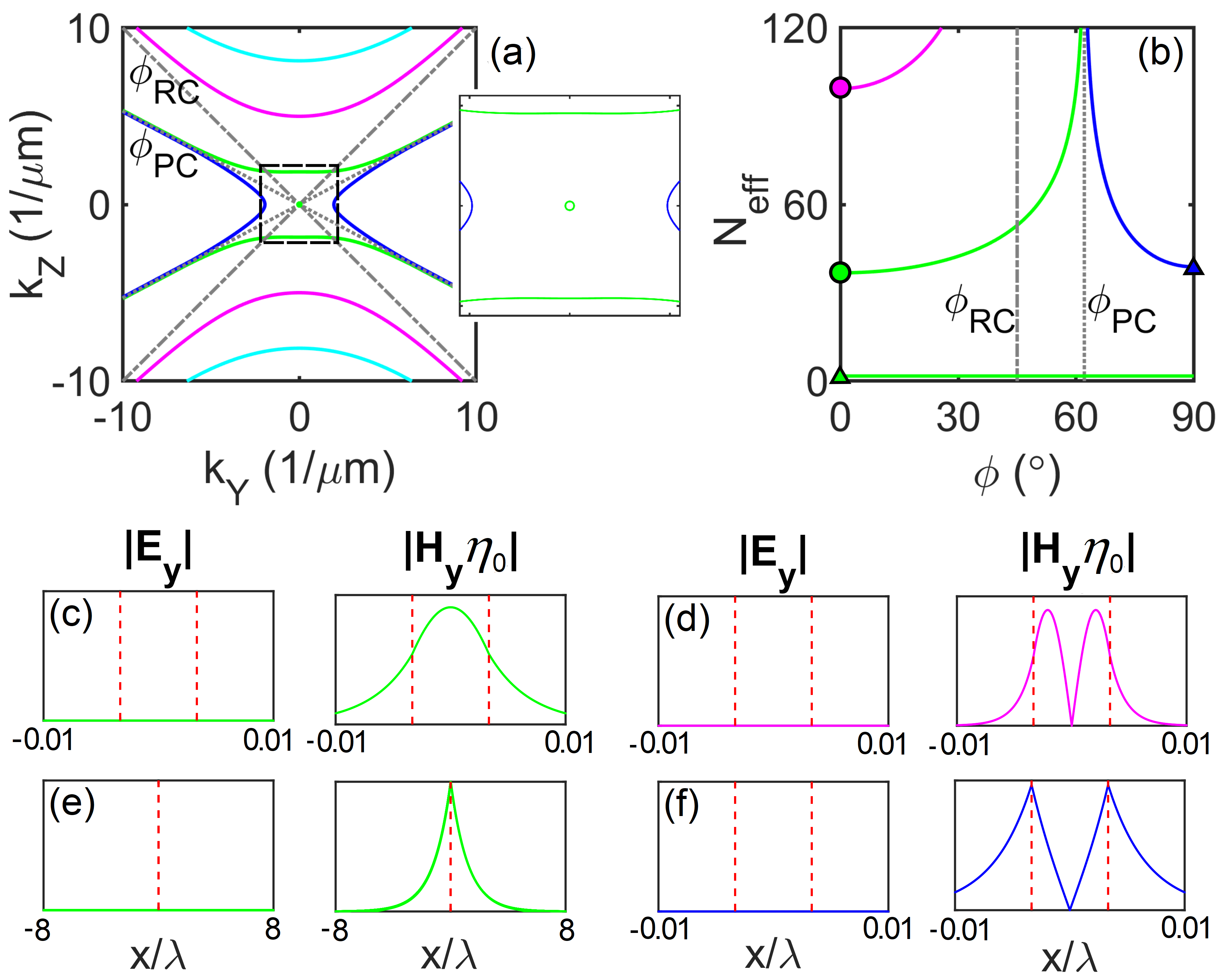}
\caption{(a) Momentum-isofrequency dispersion diagram for the modes supported by the structure (colored solid lines). The dashed and the dashed-dot lines are the hyperbolic plasmon cutoff angle $\phi_{PC}$ and the resonant cone angle $\phi_{RC}$ at the surface of a semi-infinite and a bulk HMM, respectively. The film OA is oriented along $k_Z$ in the HMM reference axis. The inset is a zoom ($K_Y, K_Z \in [-2.2, 2.2] {\mu m}^{-1}$) of the dashed squared region in (a) showing the elliptic mode. (b) Equivalent angular dispersion diagram showing $N_{eff}$ in terms of the propagation direction $\phi$ with respect to the OA. Absolute value of the field components amplitudes $E_y$ and $H_y$ showing TM dominant polarization for the marks in (b) corresponding to (c) the fundamental symmetric hyperbolic mode (green dot), (d) the first order  hyperbolic antisymmetric modes (pink dot), (e) the symmetric elliptic plasmon (green triangle) and (f) the antisymmetric hyperbolic bulk plasmon (blue triangle). Dashed lines in (c-f) are the waveguide interfaces. The waveguides parameters are $D/\lambda=0.008$, $\varepsilon_c=3$, $\varepsilon_{ef}=4$ and $\varepsilon_{of}=-4$.}
\label{Fig-2}
\end{figure}

We start our analysis by considering small electrical thicknesses, $D/\lambda$. Fig. \ref{Fig-2}(a) corresponds to the momentum isofrequency dispersion diagram for $D/\lambda=0.008$, where the film OA is oriented at $k_y = 0$, and Fig. \ref{Fig-2}(b) corresponds to the equivalent angular dispersion diagram, showing $N_{eff}$ in terms of the propagation direction, $\phi$, with respect to the OA. The two figures show the existence of different modes with hyperbolic dispersion, which contrary to a bulk HMM, exists at both, propagation near the OA and propagation orthogonal to the OA. The structure also supports one elliptic mode, [inset of Fig. \ref{Fig-2}(a)], which corresponds to a delocalized plasmon, (low $N_{eff}$), with a hybrid TM dominant polarization [Fig. \ref{Fig-2}(e)]. Near $\phi = 0^\circ$, an infinite number of hyperbolic mode branches exist. They show anomalous order  \cite{Anomalous_Ordering}, with the lower order modes having the lower effective index (see Fig. 2(c,d)). As the angle between the OA and the propagation direction is increased, these modes shows a TM dominant hybrid polarization (hTM) with $E_y \neq 0$. hTM modes are extremely localized and are weakly affected by the waveguide interfaces, and hence they reach $N_{eff} \rightarrow \infty$ at the resonance cone angle of the bulk HMM:
\begin{equation}
    \tan^2{\phi_{RC}}= \left| \varepsilon_{ef}/\varepsilon_{of} \right|,
    \label{phibulk}
\end{equation}
which in our specific system is $\phi_{RC}=45^\circ$ [dashed dotted line in Fig. \ref{Fig-2}(a,b)]. The fundamental hyperbolic mode, hTM$_0$, is however an exception, as it is less localized and it is perturbed by the waveguide interfaces. As a consequence, it reaches $N_{eff} \rightarrow \infty$ at: 
\begin{equation}
   \sin^2 \phi_{PC} = \dfrac{\varepsilon_{ef}\varepsilon_{of} - \varepsilon_c^2}{\varepsilon_{of}\left(\varepsilon_{ef}-\varepsilon_{of}\right)},
    \label{phiplasmon}
\end{equation}
which is the cutoff angle for hyperbolic-plasmon supported at the interface between semi-infinite Type II HMMs and a dielectric \cite{osamu-practical}.

\noindent At $\phi = 90^\circ$, only two modes exist: the delocalized symmetric elliptic plasmon described above, and an hyperbolic antisymmetric plasmon [dark blue line in Fig. \ref{Fig-2}(a,b)], which is localized at the film interfaces [Fig. \ref{Fig-2}(f)] and asymptotically reaches $N_{eff} \rightarrow \infty$ at $\phi_{PC}$.

\begin{figure}[ht]
\centering
\includegraphics[width=0.5\linewidth]{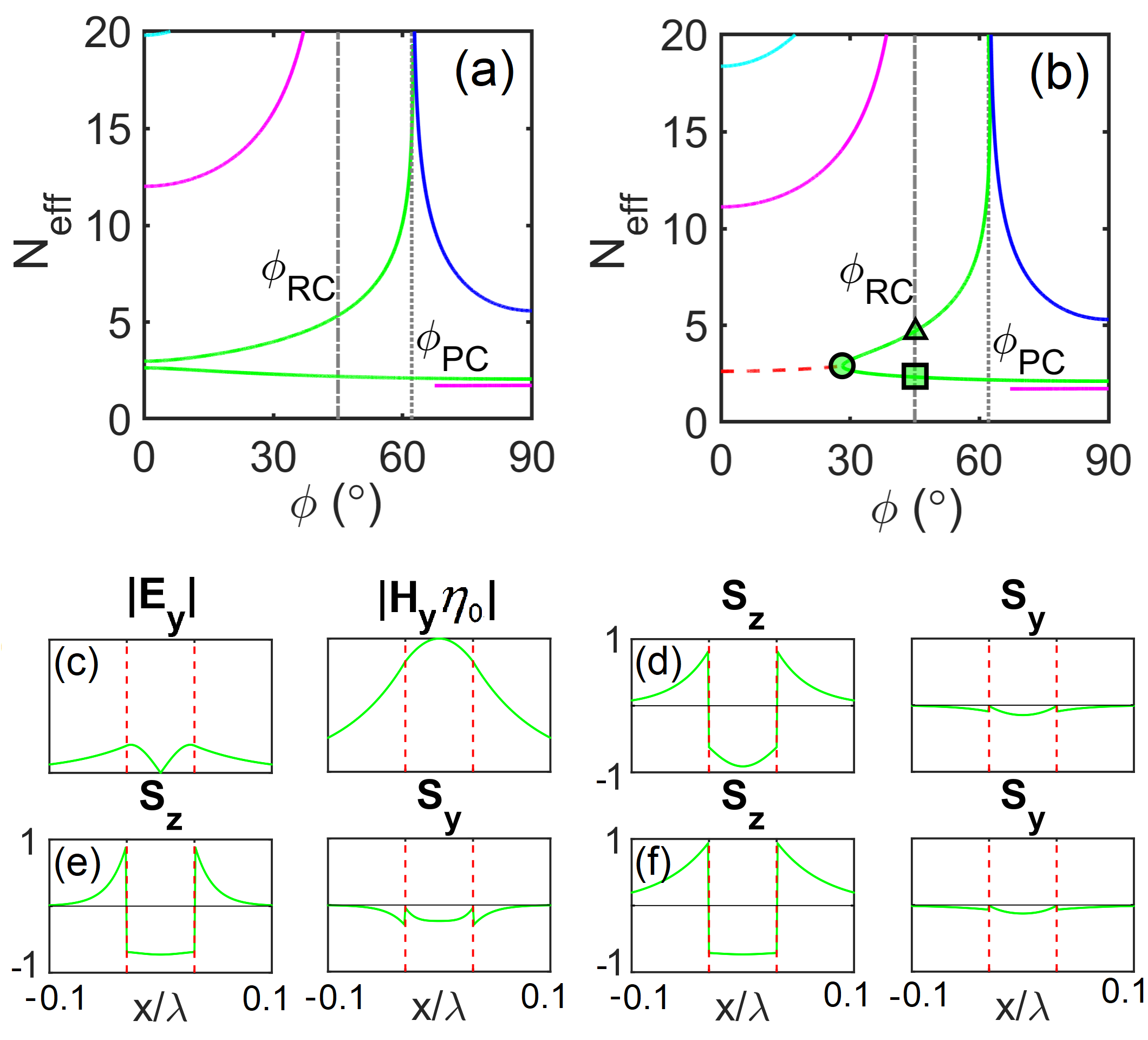}
\caption{Angular dispersion diagrams showing the first transition from (a) $D/\lambda=0.065$ to (b) $D/\lambda=0.07$. Solid color lines correspond to propagating modes while the red dashed line in (b) are Ghost waves. (c) $E_y$ and $H_y$ field amplitudes at the branch point (circle in (b)). Poynting vector components along the propagation direction, $S_z$, and the walk-off component; $S_y$, for the marks shown in (b), corresponding to (d) the branch point (slow light) at $N_{eff} \approx 2.91$ and $\phi \approx 28.25^\circ$ (circle), (e) upper branch (backward propagation) , $N_{eff} \approx 4.7$ and $\phi=45^\circ$ (triangle), (f) lower branch (forward propagation) $N_{eff} \approx 2.34$ and $\phi=45^\circ$ (square). Material parameters as in Fig. \ref{Fig-2}}
\label{Fig-3}
\end{figure}

\begin{figure}[ht]
\centering
\includegraphics[width=0.7\linewidth]{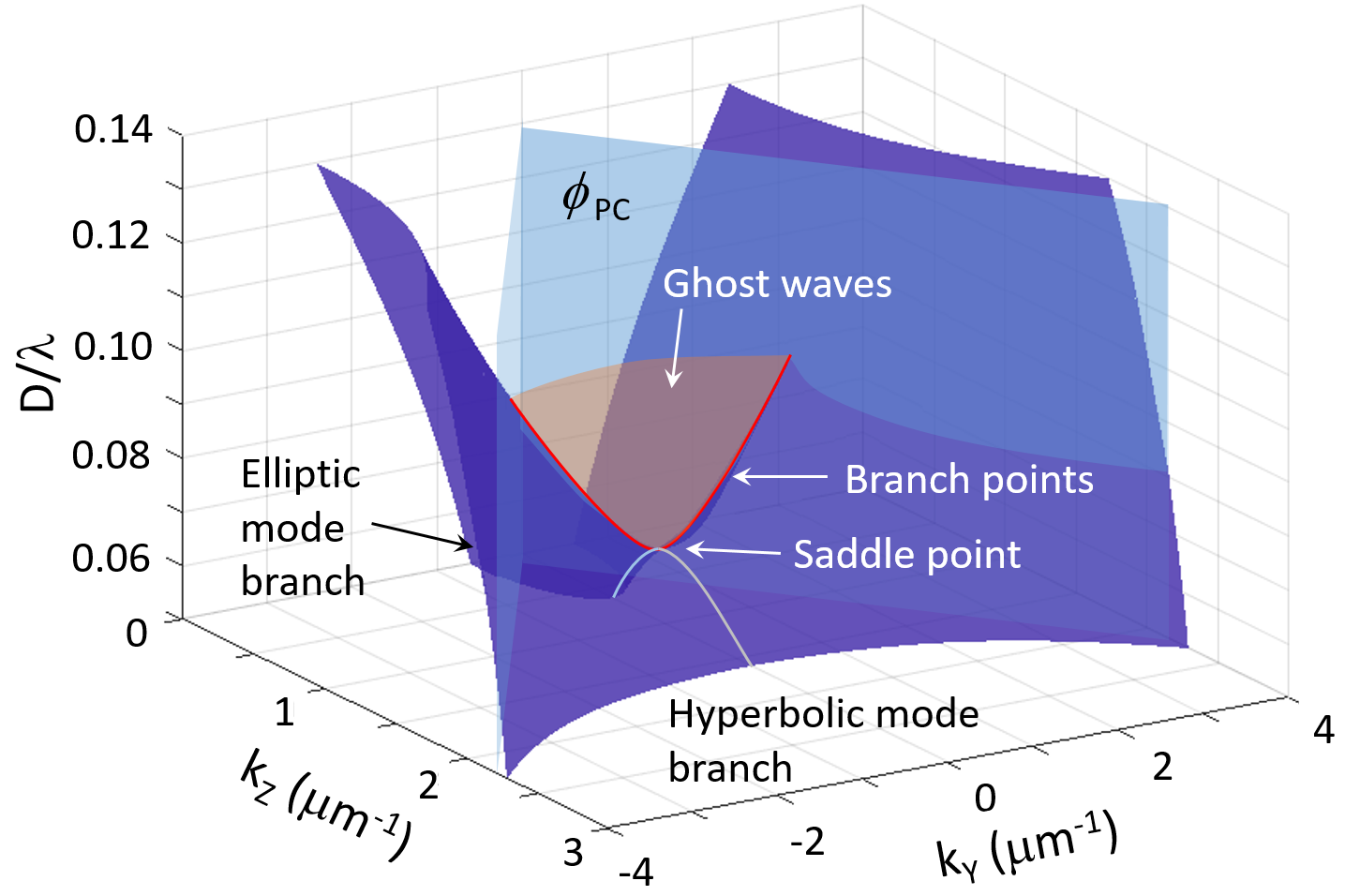}
\caption{Dispersion diagram showing $D/\lambda$ in terms of $k_Y, k_Z$. Dark blue surface is the elliptic and hyperbolic modes that merge at the saddle point. The orange surface correspond to Ghost waves. Red line correspond to the branch points. The light blue surfaces are the angle $\phi_{PC}$. }
\label{Fig-4}
\end{figure}

\noindent The results for hyperbolic meta-surfaces when $D/\lambda \longrightarrow 0$ are better described in our model by using a a biaxial HMM as a film, which changes the polarization of the delocalized elliptic symmetric mode to TE-dominint.  Therefore, in this paper we focus in the situation where $D/\lambda$ is  increased. Raising $D/\lambda$ results in a decrease of $N_{eff}$ for all the hyperbolic modes, while $N_{eff}$ increases for the symmetric elliptic plasmon. In this process, the hTM$_0$ mode and the symmetric TM plasmon effective index values start approaching for $\phi=0^\circ$. When $D/\lambda\approx 0.065$, the two modes coalesce and result in a saddle point with  $N_{eff} \approx 2.81$ [Fig. \ref{Fig-3}(a)]. As $D/\lambda$ increases, the two mode dispersion curves depart from $\phi=0^\circ$ becoming a single mode curve. The saddle point then transforms into a branch point connecting the two merged curves with a new branch of complex guided modes \cite{Jiang:09}, also known as Ghost waves \cite{narimanov-AdvPhot,PhysRevA.98.013818} [dashed red line in Fig. \ref{Fig-3}(b)]. Ghost waves feature both, oscillatory and evanescent nature along the propagation direction and hence their propagation distance is limited. As the merged branch transits to higher values of $\phi$ [Fig. \ref{Fig-3}(b)], the branch, and specifically, the branch point, shows a TM-dominant hybrid polarization [Fig. \ref{Fig-3}(c)]. At $D/\lambda \approx  0.109$, the branch point crosses the propagation direction $\phi=\phi_{PC}$, transforms into a inflection point and the Ghost branch disappears. Eventually, the merged branch of propagating modes ends up forming a hyperbolic branch of TM-dominant symmetric plasmons propagating at the proximity of $\phi = 90^\circ$. In summary, the symmetric plasmon and the fundamental hyperbolic mode have performed a topological transition, changing its nature from an elliptical/hyperbolic to an hyperbolic dispersion as the waveguide electrical thickness increases. Figure \ref{Fig-4} shows the full dispersion diagram, $D/\lambda$ in terms of $k_Y, k_Z$, illustrating the topological transition occurring at the saddle point, the branch points, the Ghost waves and the asymptotic behavior of the mode around the plane $\phi_{PC}$.

We examine the group velocity of the different modes in the structure, with $\vec{v}_g=\nabla_k \omega ( \vec{k})$, which in our case $\vec{v}_g =v_{gy} \hat{y}+v_{gz} \hat{z} \propto \partial (D/\lambda)/\partial k_Y \hat{y}+\partial (D/\lambda)/\partial k_Z \hat{z}$. According to this definition, $v_g=0$ at the saddle point in the dispersion diagram and results in slow light when the wave propagates along the optical axis, as previously reported in similar structures \cite{Slow_light,Jiang:09,tunableslow}. For any other direction, $v_g$ is orthogonal to the isofrequency dispersion lines. An special case are the branch points, where from Fig. \ref{Fig-4} we see that $\partial (D/\lambda)/\partial k_Z=0$, and therefore the group velocity is null along the propagation direction ($v_{gz}=0$), showing slow light in this direction.  

To determine the origin of the zero group velocity along the propagation direction, we examine the $S_y$ and $S_z$ components of the averaged Poynting vector across the waveguide for the three points near the branch point marked in Fig. \ref{Fig-3}(b). The results shows the typical plasmon behavior, with negative and positive power flow in the film and substrate, respectively, and a small walk-off, $S_y \neq 0$ related with $v_{gy} \neq 0$. The total power flow along the propagation direction, obtained by integrating the $S_z$ component of the Poynting vector along the $x$ direction, is null at the branch point [Fig. \ref{Fig-3}(d)], resulting in $v_{gz}=0$. In the upper branch, the power flow is dominated by the Poynting vector in the film [Fig. \ref{Fig-3}(e)], resulting in $v_{gz} < 0 $,  while in the lower branch the power flow is dominated by the substrate [Fig. \ref{Fig-3}(f)], resulting in $v_{gz} > 0$. These results are consistent with the fact that the higher the value of $N_{eff}$ in a mode, the more localized is the field in the film, so that the localization determines the direction of the power flow. Consequently, slow light at the branch point results from an exact balance between the two film and substrate counter-propagating power flux. When the branch point transforms into an inflection point at $D/\lambda > 0.109$, and the branch of Ghost waves disappears, all the new hyperbolic branch shows forward propagation, $S_z > 0$. Regarding the other modes supported by the structure, hyperbolic branches near the OA shows backward propagation, $S_z<0$, while Ghost modes, results in an null averaged pointing vector due to their dual evanescent/oscillatory nature \cite{narimanov-AdvPhot}.
\begin{figure}
\centering
\includegraphics[width=0.7\linewidth]{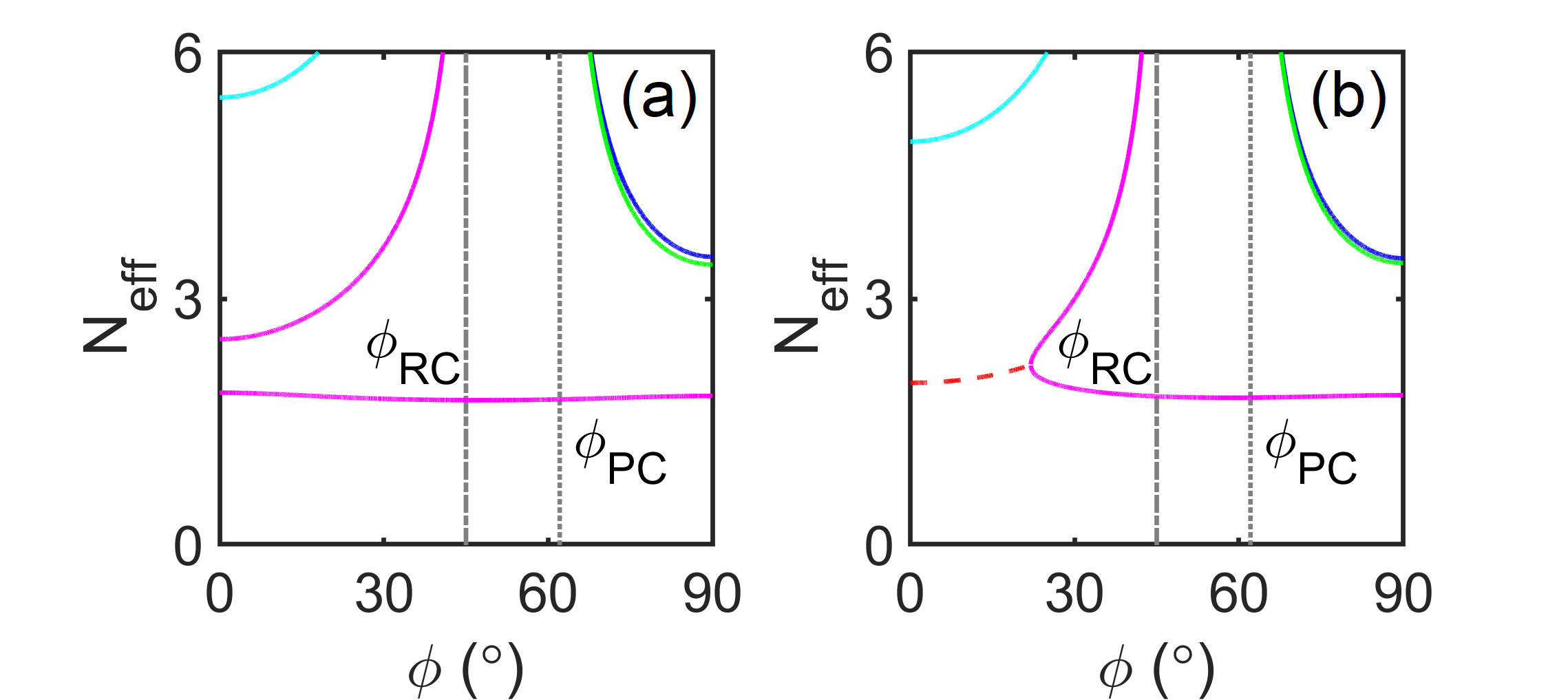}
\caption{Angular dispersion diagrams for the second transition from (a) $D/\lambda=0.22$ to (b) $D/\lambda=0.24$. Solid lines are propagating modes while the dashed red line in (b) is the Ghost wave.}
\label{Fig-5}
\end{figure}

Along with the topological transition, at the electrical thickness $D/\lambda \approx 0.0168$, a new branch of elliptic hybrid mode starts to appear near $\phi = 90^\circ$ [bottom pink line in Fig. \ref{Fig-3}(a-b)]. As we keep increasing the frequency, the elliptic branch extends to lower propagation angles and for $D/\lambda \approx 0.2$ it reaches $\phi=0^\circ$, thus the mode propagates for any direction [Fig. \ref{Fig-5}(a)]. Similar to what occurs in anti-crossings at the proximity of a Dirac point \cite{gomis-dirac}, the mode changes parity and polarization along the mode branch, so that it is a TM$_1$ mode at $\phi=0^\circ$  and a TE$_0$ mode at $\phi=90^\circ$. When $D/\lambda=0.227$ this elliptic branch coalesces with the hTM$_1$ mode at $\phi=0^\circ$ and $N=2.103$, resulting in a new saddle point which again shows slow light and a new topological transition starts. In this transition, a new branch point with a new ghost branch appears. The branch point features slow light and starts moving to higher values of $\phi$ as $D/\lambda$ increases [Fig. \ref{Fig-5}(b)]. However, in this case, the branch point will disappear after crossing the resonant cone angle, $\phi_{RC}$. At the end of the transition, the initially hybrid elliptical TM$_1$/TE$_0$ (at $\phi=0\circ/ 90^\circ$) and hyperbolic hTM$_1$ merged branches transform into an hyperbolic hybrid hTE$_0$-dominant branch that asymptotically reaches infinity at $\phi_{RC}$. 

The transition described above is repeated for higher order modes. As $D/\lambda$ increases, hybrid TM$_{n+1}$/TE$_n$ elliptical branches coalesces with anomalously ordered hyperbolic hTM$_{n+1}$ modes at $\phi=0^\circ$, forming a saddle point. This starts the topological transition where a branch of Ghost modes and a branch point showing slow light appears. The transition ends up into an hyperbolic hTE$_n$ branch in the proximity of $\phi=90^\circ$, which shows normal ordering (lower-order $n$ have higher $N_{eff}$). As a consequence, hyperbolic modes do not cut-off and instead perform topological transitions so that, near $\phi=90^\circ$, there is a finite set of normally ordered TE dominant hybrid hyperbolic modes and a symmetric plasmon mode with forward power flow that are originated from successive topological transitions of the infinite set of anomalously ordered TM dominant hyperbolic modes with backward power flow existing near $\phi=0^\circ$.

\newpage

The physical properties associated with the topological transition we predict in hyperbolic waveguides could have practical relevance in polarization/parity mode conversion or propagation direction-dependent low or high order mode filter. More interestingly, the existence of positive and negative Poynting vector near a branch point can result in simultaneous positive and negative refraction and finally, the existence of slow light and Ghost waves for a continuous set of electrical film thicknesses, or equivalently frequencies, open the door to obtain broadband or rainbow light trapping \cite{TrappedRainbow}.  

\vspace{1cm}

\noindent {\bf Funding}. Agència de Gestió d’Ajuts Universitaris i de Recerca (2017-SGR-1400); Ministerio de Economía y Competitividad (PGC2018-097035-B-I00, SEV-2015-0522); H2020 Marie Skłodowska-Curie Actions (GA665884); Fundación Cellex; Fundació Mir-Puig. National Science Foundation with a CAREER Grant No. ECCS-1749177.

\newpage


\end{document}